# Moiré-induced symmetry breaking of charge order in van der Waals heterostructures

Sandra Sajan[1,10], Laura Pätzold[2,10], Tarushi Agarwal[3], Clara Pfister[2,4], Haojie Guo[1], Sisheng Duan[1], P. V. Sruthibhai[5], Mariana Rossi[6,7], Maria N. Gastiasoro[1], Sara Barja[1,5,8,9], Ravi P. Singh[3], Tim Wehling[2,4],* and Miguel M. Ugeda[1,5,9],*

[1]*Donostia International Physics Center (DIPC), San Sebastian, Spain*
[2]*Institute of Theoretical Physics, University of Hamburg, Hamburg, Germany*
[3]*Department of Physics, Indian Institute of Science Education and Research Bhopal, Bhopal, India*
[4]*The Hamburg Centre for Ultrafast Imaging, Hamburg, Germany*
[5]*Centro de Física de Materiales (CSIC-UPV/EHU), San Sebastián, Spain.*
[6]*Max Planck Institute for the Structure and Dynamics of Matter, Hamburg, Germany*
[7]*Yusuf Hamied Department of Chemistry, University of Cambridge, Cambridge, UK*
[8]*Dept. Polymers and Advanced Materials, University of the Basque Country (UPV/EHU), San Sebastián, Spain*
[9]*Ikerbasque, Basque Foundation for Science, Bilbao, Spain*
[10]*These authors contributed equally.*

\* *Corresponding authors:* tim.wehling@uni-hamburg.de *and* mmugeda@dipc.org

*Layered materials that stack different lattice symmetries are rare in nature. Misfit layered chalcogenides, which combine square and hexagonal lattices of rocksalt monochalcogenides and transition-metal dichalcogenides, provide a platform to explore how incommensurability and explicit symmetry breaking impact collective electronic phases. Here we use low-temperature scanning tunneling microscopy/spectroscopy to probe the misfit compounds $(MS)_{1+\delta}TaS_2$ with M = Pb, Sn and track how the misfit interface reshapes the electronic ground state of the embedded $1H\text{-}TaS_2$ monolayers. High-resolution STM imaging and Fourier analysis reveal that the charge-density wave (CDW) is incommensurate and fragments into nanometer-sized domains. Strikingly, the CDW exhibits a pronounced and anisotropic response to the uniaxial moiré potential imposed by the misfit layer: its coherence lengths and ordering wavevectors become inequivalent, demonstrating a strong nonlinear coupling between the intrinsic CDW instability and the symmetry-breaking moiré field. First-principles-informed multiscale modeling shows that this reorganization arises from the combined effect of interlayer charge transfer and the spatially anisotropic energy landscape introduced by the misfit interface. In contrast, superconductivity is comparatively insensitive to the moiré, revealing a uniform, single full-gap consistent with s-wave pairing. Our results establish heterosymmetry stacking as a route to engineer correlated states in van der Waals materials.*

## Introduction

Collective electronic states are highly sensitive to underlying lattice symmetries and superlattice modulations, providing practical knobs to reshape and tune broken-symmetry phases such as charge order and superconductivity. Two-dimensional (2D) materials are especially well suited to such control, and recent progress using twisted layers[1–3] and patterned substrates[4,5] highlights how engineered periodic potentials can be used to tune correlated phases[6]. Misfit layered compounds, a family of van der Waals materials, emerge as natural platforms to explore symmetry-breaking effects on collective phases in 2D materials without strongly perturbing their underlying electronic structure. In their simplest stoichiometry, $(MX)_{1+\delta}(TX_2)$, these layered materials combine alternating stacks of transition metal dichalcogenide monolayers ($TX_2$, with T = Ti, V, Nb, Ta, Mo, W and X = S, Se) and monochalcogenide bilayers (MX, M = Sn, Pb, Bi, Sb), the latter acting as electronic decouplers[7–9]. While the presence of charge density wave (CDW)[10,11] and superconductivity[12,13,10,14–17] is established in these materials, the microscopic mechanisms and key degrees of freedom governing these ordered states remain unknown.

From a symmetry perspective, misfit materials hold the unique property of combining lattices with different symmetries. The MX layers exhibit a rocksalt, square lattice that, upon stacking, lowers the threefold rotational symmetry ($C_3$) of the 1H-TaS$_2$ layer (point group D3h). As a result, the square–hexagonal interface generates a uniaxial moiré superlattice potential that acts as an explicit symmetry-breaking field, capable of lifting the degeneracy between symmetry-related ordering wavevectors in the TMD layer. Therefore, misfit compounds offer a unique opportunity to directly determine the barely explored impact of such symmetry reduction on delicate collective phases in two dimensions.

Here we report measurements of the electronic structure of two prototypical misfit compounds, $(PbS)_{1.13}TaS_2$ and $(SnS)_{1.15}TaS_2$, using high-resolution low-temperature scanning tunneling microscopy and spectroscopy (STM/STS) and non-contact atomic force microscopy (nc-AFM). We find that the MX-TaS$_2$ interface generates a uniaxial (1D) moiré that lifts the threefold degeneracy of the CDW wavevectors of the TMD monolayer, yielding an incommensurate CDW that fragments into nanometer-scale CDW domains. Despite the incommensurate order, the wavevector intensity exhibits maxima at rational Bragg ($Q_B$) positions at $Q_B/2$ for one component, and $3Q_B/8$ and its equivalent $5Q_B/8 = -3Q_B/8+Q_B$ for the other two components. First-principles-based multiscale modelling identifies the microscopic origin of the observed CDW state as arising from the interplay between interlayer charge transfer and the potential-energy modulation imposed by the misfit moiré. In contrast, STS measurements at 0.34 K reveal a uniform single, full superconducting gap in the TaS$_2$ layers consistent with isotropic *s*-wave pairing, alongside proximitized gaps of slightly smaller magnitude in the monochalcogenide layers, evidencing a highly transparent interface and insensitivity to the moiré potential.

## Results

Our experiments were carried out on high-quality crystals of (MS)$_{1+\delta}$TaS$_2$ with M = Pb, Sn (see methods and supplementary note 1). These layered compounds comprise alternating 1H-TaS$_2$ monolayers and rocksalt MS(001) bilayers with a lattice mismatch along a high-symmetry axis. As illustrated in fig. 1a, the square MS lattice lies epitaxially on the hexagonal TaS$_2$ lattice such that one MS in-plane axis aligns with a close-packed direction of TaS$_2$, yielding a commensurate registry along *y*. Along the orthogonal *x* direction, however, the intrinsic periodicities of the square and hexagonal lattices differ ($a_{PbS}$ = 5.78 Å, $a_{SnS}$ = 5.75 Å and $a_{TaS2}$ = 3.30 Å), producing a one-dimensional (1D) incommensurability. This mismatch accumulates laterally to generate a long-wavelength moiré modulation parallel to *x*, while the layers remain structurally locked along *y*. The impact of this moiré on the properties of the collective electronic states of the TMD layers remains poorly understood, and constitutes the main goal of this work.

Figure 1b illustrates the typical morphology of our samples – (PbS)$_{1.13}$TaS$_2$ in this case – after exfoliation (see methods), showing terraces exposing adjacent PbS and TaS$_2$ terminations with a step height of ~7 Å. Figures 1c,d show atomically resolved STM images of the PbS and TaS₂ layers, respectively. On PbS, the STM images show a square lattice where both atomic species appear with comparable topographic contrast. Atomic defects, presumably vacancies, are also present in the PbS surface. On TaS$_2$, STM images exhibit the trigonal atomic registry, upon which an apparently incommensurate superlattice is overlaid, reminiscent of a CDW modulation. We found the same morphology in the SnS and TaS₂ layers of (SnS)$_{1.15}$TaS₂ (supp. note 2). Notably, this incommensurate superlattice clearly departs from the CDW orders previously reported for isolated 1H-TaS$_2$ monolayers, close to a 3×3 or a 2×2 periodicity, depending on doping[18–20]. This observation is also in contrast to the periodicity found in the isostructural analogue $(SnS)_{1.15}TaS_2$, where a similar CDW order close to a 3×3 has been reported[10,11]. This constitutes a first indication that the MS/TaS$_2$ misfit environment stabilizes a distinct, incommensurate charge modulation.

While the presence of the 1D moiré arising from the square and hexagonal lattices is not apparent in the topography images, especially in $(PbS)_{1.13}TaS_2$ (see supp. note 2 for the $(SnS)_{1.15}TaS_2$ case), its periodicity and orientation can be extracted from the Fourier transform (FFT) analysis of the images. Figure 1e shows the real-space modulation of the moiré with respect to the crystal orientations of PbS and 1H-TaS$_2$ in the regions in figs. 1c,d. The striped moiré has a wavelength of $\lambda_m = \frac{a_{PbS} \cdot a_{TaS2}}{|a_{PbS} - a_{TaS2}|} \approx 7.7$ Å ($\lambda_m \approx 7.8$ Å in $(SnS)_{1.15}TaS_2$), with a wavevector $q_m$ parallel to the incommensurate $x$ axis (i.e., along MS [100] and TaS$_2$ [10$\bar{1}$]). This moiré periodicity corresponds to three beats within a longer-wavelength envelope (~2.3 nm for $(PbS)_{1.13}TaS_2$) that arises from the lowest-order near-commensurate matching of the MS and TaS$_2$ lattice spacings. Therefore, the 1D superlattice breaks the threefold rotational symmetry of 1H-TaS$_2$, reducing it to a twofold rotational symmetry aligned with the misfit $y$ axis.

To study the microscopic properties of the electronically ordered phases in these misfit compounds, we first investigated the electronic structure of both surface terminations (PbS and TaS$_2$) using low-temperature STS. Figure 2a shows two typical d$I$/d$V$ spectra (d$I$/d$V \propto$ local density of states) acquired consecutively on two alternating terraces of PbS and TaS$_2$ (supp. Note 3 for the $(SnS)_{1.15}TaS_2$ data). The observed electronic structure of TaS$_2$ monolayer is dominated by a broad resonance located at +0.9 V with a significantly reduced density of states (DOS) at lower energies down to -1 V. The DOS lineshape is reminiscent of previous STS measurements on TaS$_2$ monolayers on graphene substrates[20,21], where mutual electronic coupling is minimal[22]. We attribute this resonance to $d$-states in the flat region of the Ta-derived band near $\bar{\Gamma}$, similar to other metallic TMD monolayers[23,24]. These observations are consistent with recent ARPES measurements[10,15,16] on $(PbS)_{1.13}TaS_2$, which reveal a negligible interlayer hybridization and a largely rigid, electron-doped TaS$_2$ band structure (~0.2 electron per Ta)[15]. In the PbS bilayer, the electronic structure shows a reduced DOS in the broad range from +0.1 V to -1 V. This agrees qualitatively with a predicted gap of ~0.2 eV for isolated PbS together with a ~1 eV depletion of the DOS around $E_F$ due to a steep band dispersion around $\bar{L}$ (ref. [25]). Experimentally, however, the PbS bilayer shows a reduced but finite DOS in the STS measurements, a behavior also expected in the PbS layers when embedded in the misfit compound that implies a charge transfer to the TaS$_2$ layer[25].

The metallic character of both PbS and 1H-TaS$_2$ layers enables the development of superconductivity in each layer, which we probed using high-resolution STS at 0.34 K. While the presence of superconductivity in the Pb- and Sn-based (MS)$_{1+\delta}$TaS$_2$ compounds is established primarily via mesoscopic techniques[10,12,14,15,17,26] (supp. note 1), their microscopic, local properties remain scarcely explored[11]. Previous transport measurements in these and related compounds indicate a robust 2D Ising character with a large, anisotropic $H_{C2}$ consistent with 2D Tinkham scaling[10,14,15,17,26,27]. Figure 2b shows four representative d$I$/d$V$ spectra acquired consecutively on the TaS$_2$ monolayer and the corresponding monochalcogenide layer in the (PbS)$_{1.13}$TaS$_2$ and (SnS)$_{1.15}$TaS$_2$ compounds. As can be seen, all layers display a hard gap with coherence peaks characteristic of superconductivity. In TaS$_2$ layers, we consistently observe for both compounds a single gap that is well described by an isotropic BCS fit, as shown in fig. 2c. The BCS gap values – $\Delta_{TaS2}$ = 0.38 ± 0.03 meV in (PbS)$_{1.13}$TaS$_2$ and $\Delta_{TaS2}$ = 0.33 ± 0.03 meV in (SnS)$_{1.15}$TaS$_2$ – together with the critical temperatures of these compounds ($T_C$ = 2.7-3 K), yield 2$\Delta/k_BT_C$ ratios of 2.6-3.3, which lie slightly below the weak-coupling BCS value (3.53).

In PbS and SnS, the superconducting gaps are slightly smaller than those in TaS$_2$, consistent with a proximitized superconductivity in the monochalcogenide layers. In both cases, the gaps are well fit by a BCS form (supp. note 4). Furthermore, both the monochalcogenide and dichalcogenide layers in (PbS)$_{1.13}$TaS$_2$ show similar responses to out-of-plane magnetic fields ($H^\perp$), with a common critical value of $H^\perp_{C2} \simeq 0.2$ T (supp. note 5). The large gap ratios $r = \Delta_{PbS,SnS}/\Delta_{TaS2}$ near unity point to strong interlayer coupling and a highly transparent interface for superconductivity[28,29]. Lastly, spatially resolved STS data shows uniform gaps on both the TMD and monochalcogen layers, which indicates strong hybridization and resultant proximity coupling between the layers. However, the superconductivity appears to be insensitive to the moiré potential within our resolution (supp. note 6). Overall, our observation of a single, BCS-like superconducting gap in the monolayers of TaS$_2$ and PbS(SnS) contrasts with reports of unconventional pairing phenomenology, including nodal[20] and mixed singlet–triplet[11] gap structures.

We next focus on probing the CDW order in the 1H-TaS$_2$ monolayers. As mentioned above, the STM images of TaS$_2$ show a 2D superlattice that appears incommensurate with the underlying lattice (fig. 1d). While compatible with CDW order, this incommensuration strikingly contrasts with the nearly commensurate CDW order reported for 1H-TaS$_2$ in bulk crystals and down to the monolayer limit, including in misfit compounds[10,11,27]. To better understand and interpret the observed incommensurate 2D superlattice, we perform Fourier analysis of atomically resolved, STM topographies of TaS$_2$ across extended fields of view. Figure 3a shows a large-scale STM image of TaS$_2$ in (PbS)$_{1.13}$TaS$_2$ with atomic resolution, whose FFT is shown in fig.3b. Apart from the Bragg peaks $\mathbf{Q}_B$, the FFT reveals three elongated features labelled as $q_i$ ($i$ = 1,2,3) along the Bragg directions and a sharp peak at $\mathbf{q}_{moiré}$ = 0.82 Å$^{-1}$·$\mathbf{q}_x$. The latter corresponds to a 1D periodicity in real space of $\lambda = 2\pi/(0.82$ Å$^{-1}) \simeq 7.7$ Å, which we attribute to the striped moiré with wavelength $\lambda_m$ formed by the square and hexagonal lattices. The elongated features $q_i$, aligned with the crystal TaS$_2$ lattice, exhibit two distinct internal structures. The $q_1$ feature, orthogonal to $\mathbf{q}_{moiré}$, shows intensity along an extended region around $\frac{1}{2}Q_B$ (2×2 in real space), where the intensity is maximum (fig. 3c). Instead, $q_2$ and $q_3$, rotated 30° with respect to $\mathbf{q}_{moiré}$, share an equal distribution of intensities with two maxima at $\frac{3}{8}Q_B$ and $\frac{5}{8}Q_B$ and a slow decay toward $\frac{1}{2}Q_B$ – the $\overline{M}$ point (fig. 3c). The shape and maxima of these features are independent of the STM imaging bias voltage (fig. 3d), which rules them out as quasiparticle interference features and are, instead, fully compatible with CDW order. The intensity profiles along $q_1$ and $q_{2,3}$ shown in fig. 3c reveal nearly equal extents of the lines around $\overline{M}$ despite the maxima occurring at different wavevectors. Remarkably, no noticeable intensity has been found close to $\frac{1}{3}Q_B$, which corresponds to the usual near 3×3 CDW periodicity in real space for TaS$_2$. We found similar behavior of the CDW phase in the TaS$_2$ layers of (SnS)$_{1.15}$TaS$_2$ compound (supp. note 7).

To gain knowledge on the CDW order in the TaS$_2$ layers embedded in misfit compounds, we visualize the $q_i$ features in real space by taking the inverse Fourier transform of the modulus of each component. Figure 3e shows the resulting real-space maps of each $q_i$ component (blue panels) within the region boxed in fig.3a along with the 1D-moiré oscillation as a reference. In the maps, light-blue regions indicate coherent CDW oscillations within a domain whereas dark regions correspond to vanishing CDW amplitudes at the domain boundaries. As seen, all three $q_i$ components are inhomogeneous and exhibit multiple nanometer-sized domains. The presence of CDW domains induce the splitting of the CDW peaks by $\Delta q_i = 2\pi/L_i$, where $L$ is the length of the domain[30]. Multiple size domains within the analyzed regions lead to size-varying splits, which ultimately form a continuous, elongated stripe feature in the FFT. In the case of $q_2$ and $q_3$, since most of the intensity is located around the two maxima at $\frac{3}{8}Q_B$ and $\frac{5}{8}Q_B$, we can take $\Delta q_{2,3} \simeq \frac{1}{4}Q_B =$ 0.475 Å$^{-1}$, implying short domains of lengths $L_{2,3} \simeq 13$ Å. For $q_1$, the signal is spread around a long arc centered on $\overline{M}$ implying a wide range of domain sizes. Nevertheless, a value of $\Delta q_{1,FWHM} \simeq 0.14$ Å$^{-1}$ from the profile of fig. 3c provides a lower bound on the y-direction domain size, giving $L_1 \simeq$ 45 Å, in good agreement with the typical longer domain lengths shown in fig. 3e for $q_1$.

These observations demonstrate the full incommensuration of the CDW order in the 1H-TaS$_2$ monolayers of the (MS)$_{1+\delta}$TaS$_2$ (M = Pb, Sn) compounds. This behavior explains why previous experiments, such as resistivity and Raman measurements exhibit no CDW-related anomalies[15–17,27], and ARPES detects no band folding, consistent with suppression of a commensurate CDW. This moiré-induced exotic CDW order departs from the nearly commensurate and $C_3$-symmetric behavior reported for this TMD metal.

To better understand and interpret our experimental findings on the threefold degeneracy breaking in the CDW order of the monolayer TMD, we performed first-principles based simulations of the CDW thermodynamics using monolayer 1H-TaS$_2$ as a minimal model system. Employing downfolding-based structural relaxations and replica-exchange molecular dynamics (REMD), we simulated an 18×18 supercell. Starting from pristine 1H-TaS$_2$, the structure relaxes into the 3×3 CDW reconstruction for the undoped system (see Supp. note 8 for the model relaxation). For comparison with the experiments, we analyze the structure factors corresponding to the relaxed structures obtained via REMD simulations in fig. 4 (see Supp. note 8 for its definition), corresponding to the Fourier-transformed STM topographs. In the pristine case in fig. 4a, the dominant spectral weight appears at $\frac{1}{3}Q_B$, corresponding to the 3×3 CDW reconstruction. Upon electron doping, our model relaxations reveal multiple competing commensurate and incommensurate CDW states (see phase diagram in Supp. note 9), with the dominant CDW instability vector shifting progressively towards the Brillouin zone edge. For an electron doping of 0.2e per TaS$_2$ unit cell, the thermodynamically relevant CDW peaks are shown in Fig. 4c. Rather than sharp peaks as in the pristine case, we observe an elongated region of enhanced intensity centered around $\frac{4}{9}Q_B$ that extends towards the zone boundary at $\frac{1}{2}Q_B$. These broadened intensity features are consistent with the short CDW coherence lengths observed experimentally. Doping thus explains the shifting of the CDW vectors away from $\frac{1}{3}Q_B$ (refs. [7,27,31]) as well as the short coherence lengths, though not the threefold CDW degeneracy breaking seen in the experiments.

A key distinction between the misfit compound and the monolayer model considered so far is the presence of the one-dimensional moiré superstructure. To assess its impact on the emergent CDW state, we incorporate hybridization effects between 1H-TaS$_2$ and the underlying MS layer through a spatially modulated electronic potential acting on the Wannier orbitals of the downfolded Hamiltonian. Specifically, we impose a vertically striped pattern with moiré periodicity and an amplitude of 0.01 Ry (0.136 eV). For the undoped model (fig. 4b), the moiré potential lifts the degeneracy of the $\frac{1}{3}Q_B$ peaks through a change in their intensities. In addition, a smeared contribution emerges around $q_{moiré}$. Doping enhances the response of the CDW state to the moiré potential. The doped system with moiré potential (fig. 4d) shows a stronger distinction between those CDW Fourier spots orthogonal and non-orthogonal to $q_{moiré}$, and reveals distinct spreads in k-space between these two sets of CDW peaks.

Our minimal model captures essential ingredients of the misfit heterostructure while deliberately reducing its microscopic complexity. As expected for such a simplified description, certain details differ from experiment: in the model, the moiré-induced symmetry breaking manifests as variations in peak intensities and k-space extent rather than as distinct maxima at shifted wave vectors. Notwithstanding these differences, the calculations show that the combination of a doping-induced shift towards a 2×2–like instability and an enhanced susceptibility to the moiré modulation (Fig. 4d) lead to a markedly strong symmetry breaking of the CDW, the shifts of the CDW maxima towards the Brillouin zone boundary and the reduced CDW coherence at a qualitative level (cf. fig. 3b).

## Discussion

Our results reveal a clear hierarchy in the sensitivity of collective electronic phases to the misfit interface. Superconductivity remains fully gapped, Ising type and well described by an isotropic BCS spectrum in the 1H-TaS$_2$ layers, with slightly reduced proximity-induced gaps in the adjacent monochalcogenide layers, consistent with a highly transparent interface. Our experiments do not show any signatures of multigap[11,14,15] or unconventional pairing[11,20], and the superconductivity appears comparatively insensitive to the symmetry breaking imposed by the misfit structure.

In contrast, the charge-density wave proves to be markedly more susceptible. While CDW order persists along the reciprocal-lattice directions defined by the 1H-TaS$_2$ Bragg vectors $Q_B$, its internal degrees of freedom are strongly reshaped: the magnitudes of the CDW wave vectors, the CDW coherence lengths, and the relative Fourier intensities along the three inequivalent $Q_B$-directions become markedly unequal. Because the moiré wave vector lies along a distinct reciprocal-space direction, $q_{moiré} \nparallel Q_B$, this behavior cannot arise from simple locking, but instead reflects nonlinear, anharmonic coupling between the misfit-induced potential landscape and a near-degenerate manifold of CDW instabilities in the $Q_B$-directions. First-principles-informed modeling shows that interlayer charge transfer shifts the CDW Fourier peaks towards the zone boundary, while the moiré-modulated interlayer potential lifts the residual three-fold rotational degeneracy, stabilizing an anisotropically deformed CDW state. These findings demonstrate that CDW formation in 1H-TaS$_2$ is robust in its preferred ordering directions, yet intrinsically soft in its wave-vector magnitude and coherence.

Taken together, our work demonstrates the viability of imposing superlattice potentials on 2D materials to selectively manipulate delicate collective electronic phases while leaving their underlying electronic structure largely intact, as exemplified by the misfit compounds studied here. More broadly, symmetry-mismatched stacking offers an accessible route to engineer such superlattice potentials and to ultimately manipulate correlated phases in van der Waals materials.

## Methods

*Crystal synthesis:* Single crystals were grown by the chemical vapor transport (CVT) method. Pre-reacted polycrystalline powders were used as the source material for the synthesis of single crystals. The polycrystalline precursors were prepared by thoroughly mixing the constituent elements in their respective stoichiometric ratios and heating the mixtures at 900°C. The resulting pre-reacted powders were sealed in evacuated quartz ampoules together with transport agents. Iodine served as the transport agent for $(PbS)_{1.13}TaS_2$, while $NH_4Cl$ was used for $(SnS)_{1.15}TaS_2$. The sealed tubes were then placed in a two-zone tubular furnace with the growth zone temperature set to 850°C for $(PbS)_{1.13}TaS_2$ and 800°C for $(SnS)_{1.15}TaS_2$, and maintained for 10 days. Single crystals of few millimeters in size were obtained for both compounds.

*STM/STS and nc-AFM measurements:* All experiments were performed in two independent ultra-high vacuum (UHV) systems (Unisoku Co., LTD): (I) an USM-1300 that operates at temperatures down to 0.34 K and equipped with perpendicular magnetic fields up to 11T and (II) a closed-cycle cryogen-free STM/nc-AFM combined microscope (USM-1800) operating at 3.8 K. In both systems, single crystals of $(PbS)_{1.13}TaS_2$ and $(SnS)_{1.15}TaS_2$ were mechanically exfoliated in UHV conditions. Afterwards, the sample was directly transferred to the STM/nc-AFM stage. Data were acquired at a temperature detailed in each figure caption. Tunneling spectroscopy (d$I$/d$V$) data were recorded using standard lock-in amplifier techniques, where a modulation voltage ($V_{a.c.}$) at a frequency of 833 Hz is coupled to the sample bias voltage ($V_s$). We used typically $V_{a.c.}$ values of 5-20 µV for d$I$/d$V$ spectra within meV range and 1-5 mV for larger energy scales (± 1 V). For constant-height nc-AFM imaging, a tuning-fork-based qPlus sensor[32] (resonant frequency $f_0 \approx 28.7$ kHz; Q = 28855) was employed. The microscope was operated in frequency-shift mode while maintaining a constant oscillation amplitude of 12 Å. Before each experiment, the STM/nc-AFM tips (made of Pt/Ir) were metallized on single crystals Au(111) or Cu(111) and calibrated against their corresponding Shockley surface state. STM/STS and nc-AFM data analyzing and rendering were carried out with the WSxM software[33].

*Density functional theory and downfolding model:* DFT calculations were performed via QUANTUM ESPRESSO[34] using the PBE approximation for the exchange-correlation potential[35]. For 1H-TaS$_2$, norm-conserving pseudo-potentials from the pslibrary database[36,37] were utilized. Plane waves until an energy cutoff of 100 Ry were included and a Fermi-Dirac-type smearing of 0.001 Ry was imposed. We included a vacuum layer of approximately 12 Å above the monolayer and used an 18×18 **k**-point grid for the unit cell. Relaxation of the unit cell for fixed cell height yielded a lattice constant of 3.34 Å for the undoped model at $n = 1.0$, with 3.35-3.41 Å for the doped models at $n$ = 1.05-1.3. The DFPT calculations were also done via QUANTUM ESPRESSO, where we used a 6×6 **q**-point grid. The generation of maximally localized Wannier functions was performed with the WANNIER90 code[38] as interfaced via QUANTUM ESPRESSO, with three energy bands around the Fermi energy transformed into the Wannier basis starting from orbitals of Ta $d_z^2$-, $d_{x^2-y^2}$, and $d_{xy}$-character. The electron-phonon coupling was determined via the EPW code[39]. The relaxations on large supercells (18×18) were facilitated by the "model III" downfolding scheme from Ref. [40], which has been implemented in the ELPHMOD package[41]. Their equivalence to DFT had been established for a variety of metallic TMD materials[40]. For the model setup, the same parameters as in DFT were used except for the **k**-point grid, which corresponds to a $\Gamma$-only calculation on the 18×18 supercell.

*Molecular dynamics simulations:* A series of MD calculations were performed for 18×18 supercells of 1H-TaS$_2$ in order to model the thermodynamic stability of the CDW configuration. For temperatures of (5, 10, 20, 50, 100) K, single-ensemble NVT simulations were carried out using the i-PI code[42–44]. The supercells were simulated over ~7 ps to thermalize the system, using a simulation time step size of 4 fs. Canonical sampling was enforced by using the stochastic-velocity-rescaling thermostat method[45] with a relaxation time parameter of $\tau$ = 800 fs. NVT simulations of 90 replicas between 5 K and 100 K were performed using the replica-exchange MD method[46] as implemented in i-PI, with the simulation duration of the system spanning ~110 ps (simulation time step size: 4 fs). To ensure canonical sampling, the PILE-G thermostat method[47,48] was used, with a relaxation time parameter of $\tau$ = 100 fs. Due to the computational complexity of the performed simulations, our data shows large fluctuations below approx. 20K, while data above that value converges sufficiently. Thus, we depict the averaged intensities at T=20K for comparison with experiment.

# References

1. Cao, Y. *et al.* Correlated insulator behaviour at half-filling in magic-angle graphene superlattices. *Nature* **556**, 80–84 (2018).

2. Cao, Y. *et al.* Unconventional superconductivity in magic-angle graphene superlattices. *Nature* **556**, 43–50 (2018).

3. Andrei, E. Y. *et al.* The marvels of moiré materials. *Nat. Rev. Mater.* **6**, 201–206 (2021).

4. Forsythe, C. *et al.* Band structure engineering of 2D materials using patterned dielectric superlattices. *Nat. Nanotechnol.* **13**, 566–571 (2018).

5. Cai, X. *et al.* Stack of correlated insulating states in bilayer graphene kagome superlattice. Preprint at https://doi.org/10.48550/arXiv.2602.16210 (2026).

6. Mellado, P., Muñoz, F. & Cabezas-Escares, J. Excitonic Charge Density Waves in Moiré Ladders. Preprint at https://doi.org/10.48550/arXiv.2512.05696 (2025).

7. Leriche, R. T. *et al.* Misfit Layer Compounds: A Platform for Heavily Doped 2D Transition Metal Dichalcogenides. *Adv. Funct. Mater.* **31**, 2007706 (2021).

8. Ng, N. & McQueen, T. M. Misfit layered compounds: Unique, tunable heterostructured materials with untapped properties. *APL Mater.* **10**, 100901 (2022).

9. Zhong, H. *et al.* Revealing the two-dimensional electronic structure and anisotropic superconductivity in a natural van der Waals superlattice $(PbSe)_{1.14}NbSe_2$. *Phys. Rev. Mater.* **7**, L041801 (2023).

10. Li, Z. *et al.* Beyond Conventional Charge Density Wave for Strongly Enhanced 2D Superconductivity in $1H-TaS_2$ Superlattices. *Adv. Mater.* **36**, 2312341 (2024).

11. Almoalem, A. *et al.* Mixed Triplet-Singlet Order Parameter in Decoupled Superconducting 1H Monolayers of Transition-Metal Dichalcogenides. Preprint at https://doi.org/10.48550/arXiv.2509.13303 (2025).

12. Sankar, R. *et al.* Superconductivity in a Misfit Layered $(SnS)_{1.15}(TaS_2)$ Compound. *Chem. Mater.* **30**, 1373–1378 (2018).

13. Yang, X. *et al.* Superconductivity in a misfit compound $(PbSe)_{1.12}(TaSe_2)$. *Supercond. Sci. Technol.* **31**, 125010 (2018).

14. Agarwal, T. *et al.* Anomalous magnetotransport in the superconducting architecturally misfit layered system $(PbS)_{1.13}TaS_2$. *Phys. Rev. B* **112**, 014501 (2025).

15. K. P., S. *et al.* Ising superconductivity in the bulk incommensurate layered material $(PbS)_{1.13}(TaS_2)$. *Phys. Rev. B* **111**, 054509 (2025).

16. Sun, X. *et al.* Tunable Mirror-Symmetric Type-III Ising Superconductivity in Atomically-Thin Natural Van der Waals Heterostructures. *Adv. Mater.* **37**, 2411655 (2025).

17. Wang, J. *et al.* Transport and thermoelectric signatures of Ising superconductivity and charge density wave in the misfit layered compound $(SnS)_{1.15}TaS_2$. *J. Phys. Condens. Matter* **37**, 255601 (2025).

18. Lin, H. *et al.* Growth of atomically thick transition metal sulfide films on graphene/6H-SiC(0001) by molecular beam epitaxy. *Nano Res.* **11**, 4722–4727 (2018).

19. Hall, J. *et al.* Environmental Control of Charge Density Wave Order in Monolayer 2H-TaS$_2$. *ACS Nano* **13**, 10210–10220 (2019).

20. Vaňo, V. *et al.* Evidence of Nodal Superconductivity in Monolayer 1H-TaS$_2$ with Hidden Order Fluctuations. *Adv. Mater.* **35**, 2305409 (2023).

21. Knispel, T. *et al.* Unconventional Charge-Density-Wave Gap in Monolayer NbS$_2$. *Nano Lett.* **24**, 1045–1051 (2024).

22. Dreher, P. *et al.* Proximity Effects on the Charge Density Wave Order and Superconductivity in Single-Layer NbSe$_2$. *ACS Nano* **15**, 19430–19438 (2021).

23. Ugeda, M. M. *et al.* Characterization of collective ground states in single-layer NbSe$_2$. *Nat. Phys.* **12**, 92–97 (2016).

24. Ryu, H. *et al.* Persistent Charge-Density-Wave Order in Single-Layer TaSe$_2$. *Nano Lett.* **18**, 689–694 (2018).

25. Kabliman, E., Blaha, P. & Schwarz, K. Ab initio study of stabilization of the misfit layer compound (Pb)$_{1.14}$TaS$_2$. *Phys. Rev. B* **82**, 125308 (2010).

26. Itahashi, Y. M. *et al.* Misfit layered superconductor (PbSe)$_{1.14}$(NbSe$_2$)$_3$ with possible layer-selective FFLO state. *Nat. Commun.* **16**, 7022 (2025).

27. Zullo, L. *et al.* Charge density wave collapse of NbSe$_2$ in the (LaSe)$_{1.14}$(NbSe$_2$}$_2$ misfit layer compound. *Phys. Rev. B* **110**, 075430 (2024).

28. McMillan, W. L. Tunneling Model of the Superconducting Proximity Effect. *Phys. Rev.* **175**, 537–542 (1968).

29. Blonder, G. E., Tinkham, M. & Klapwijk, T. M. Transition from metallic to tunneling regimes in superconducting microconstrictions: Excess current, charge imbalance, and supercurrent conversion. *Phys. Rev. B* **25**, 4515–4532 (1982).

30. Yan, S. *et al.* Influence of Domain Walls in the Incommensurate Charge Density Wave State of Cu Intercalated 1T-TiSe$_2$. *Phys. Rev. Lett.* **118**, 106405 (2017).

31. Niedzielski, D. *et al.* Unmasking Charge Transfer in the Misfits: ARPES and Ab Initio Prediction of Electronic Structure in Layered Incommensurate Systems without Artificial Strain. *Phys. Rev. Lett.* **135**, 206202 (2025).

32. Giessibl, F. J. The qPlus sensor, a powerful core for the atomic force microscope. *Rev. Sci. Instrum.* **90**, 011101 (2019).

33. Horcas, I. *et al.* WSXM: A software for scanning probe microscopy and a tool for nanotechnology. *Rev. Sci. Instrum.* **78**, 013705 (2007).

34. Giannozzi, P. *et al.* QUANTUM ESPRESSO: a modular and open-source software project for quantum simulations of materials. *J. Phys. Condens. Matter* **21**, 395502 (2009).

35. Perdew, J. P., Burke, K. & Ernzerhof, M. Generalized Gradient Approximation Made Simple. *Phys. Rev. Lett.* **77**, 3865–3868 (1996).

36. Hamann, D. R. Optimized norm-conserving Vanderbilt pseudopotentials. *Phys. Rev. B* **88**, 085117 (2013).

37. Dal Corso, A. Pseudopotentials periodic table: From H to Pu. *Comput. Mater. Sci.* **95**, 337–350 (2014).

38. Pizzi, G. *et al.* Wannier90 as a community code: new features and applications. *J. Phys. Condens. Matter* **32**, 165902 (2020).

39. Poncé, S., Margine, E. R., Verdi, C. & Giustino, F. EPW: Electron–phonon coupling, transport and superconducting properties using maximally localized Wannier functions. *Comput. Phys. Commun.* **209**, 116–133 (2016).

40. Schobert, A. *et al.* Ab initio electron-lattice downfolding: Potential energy landscapes, anharmonicity, and molecular dynamics in charge density wave materials. *SciPost Phys.* **16**, 046 (2024).

41. Berges, J., Schobert, A., van Loon, E. G. C. P., Rösner, M. & Wehling, T. O. elphmod: Python modules for electron-phonon models. Zenodo https://doi.org/10.5281/zenodo.17702034 (2025).

42. Ceriotti, M., More, J. & Manolopoulos, D. E. i-PI: A Python interface for ab initio path integral molecular dynamics simulations. *Comput. Phys. Commun.* **185**, 1019–1026 (2014).

43. Kapil, V. *et al.* i-PI 2.0: A universal force engine for advanced molecular simulations. *Comput. Phys. Commun.* **236**, 214–223 (2019).

44. Litman, Y. *et al.* i-PI 3.0: A flexible and efficient framework for advanced atomistic simulations. *J. Chem. Phys.* **161**, 062504 (2024).

45. Bussi, G., Donadio, D. & Parrinello, M. Canonical sampling through velocity rescaling. *J. Chem. Phys.* **126**, 014101 (2007).

46. Sugita, Y. & Okamoto, Y. Replica-exchange molecular dynamics method for protein folding. *Chem. Phys. Lett.* **314**, 141–151 (1999).

47. Ceriotti, M., Parrinello, M., Markland, T. E. & Manolopoulos, D. E. Efficient stochastic thermostatting of path integral molecular dynamics. *J. Chem. Phys.* **133**, 124104 (2010).

48. Ceriotti, M., Bussi, G. & Parrinello, M. Langevin Equation with Colored Noise for Constant-Temperature Molecular Dynamics Simulations. *Phys. Rev. Lett.* **102**, 020601 (2009).

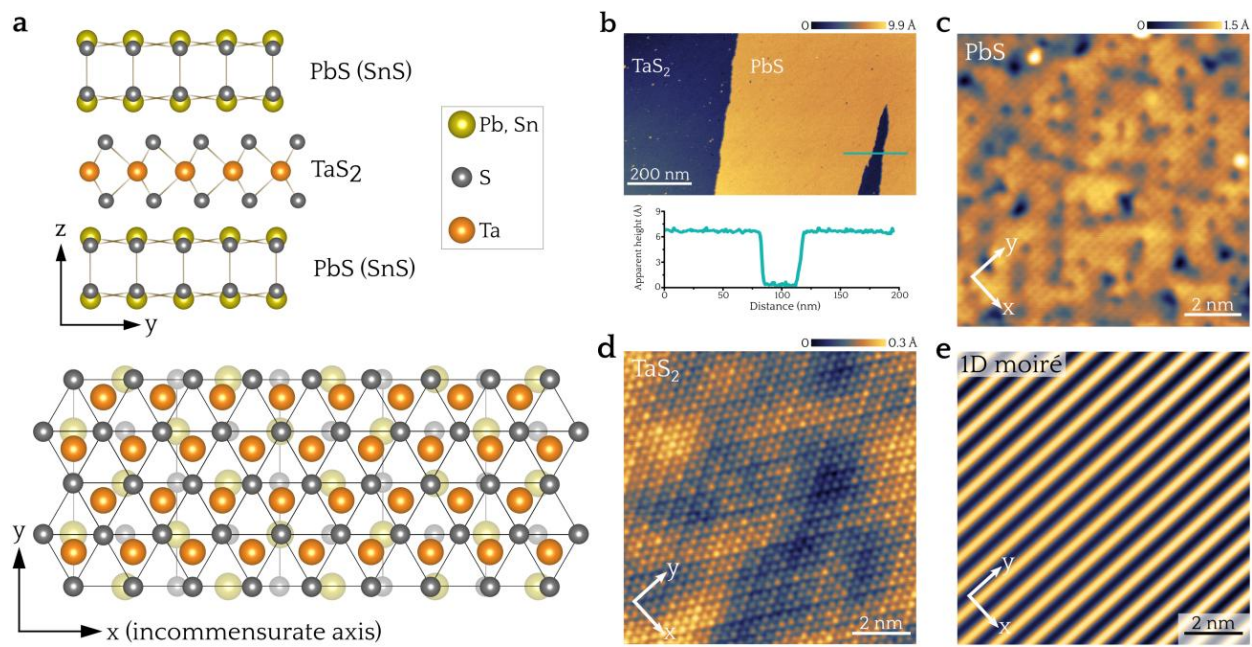

**Figure 1. Structure and moiré of monochalcogenide–TaS$_2$ misfit compounds. a,** Side and top view sketches of (MS)$_{1+\delta}$TaS$_2$ (M = Pb, Sn). **b,** Large-scale STM topography showing the PbS and TaS$_2$ terrace layers ($V_s$ = 1 V, $I$ = 100 pA, $T$ = 4.2 K). The height profile below is taken along the green line in the image. **c-d,** Atomically resolved STM images of TaS$_2$ ($V_s$ = 0.3 V, $I$ = 260 pA, $T$ = 0.34 K) and PbS ($V_s$ = 0.1 V, $I$ = 320 pA, $T$ = 4.2 K), respectively. **e,** Wavefront of the 1D moiré formed between the PbS and TaS$_2$ crystal lattices extracted by doing the inverse FFT of the image in **c**.

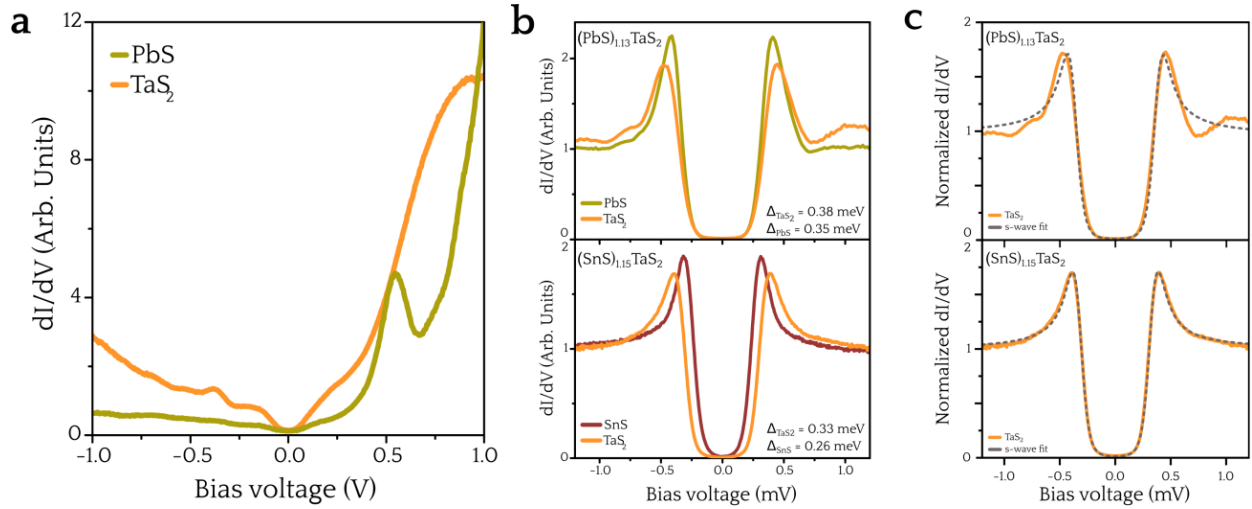

**Figure 2. Electronic structure and superconductivity. a**, Wide-bias d$I$/d$V$ spectra consecutively acquired on TaS$_2$ (orange) and PbS (green) ($T$ = 4.2 K). **b**, Low-bias d$I$/d$V$ spectra consecutively acquired on TaS$_2$ and PbS (upper panel) and TaS$_2$ and SnS (lower panel) showing a single, full superconducting gap in all cases ($T$ = 0.34 K). **c**, BCS s-wave fit to the d$I$/d$V$ spectra on TaS$_2$ shown in **b**.

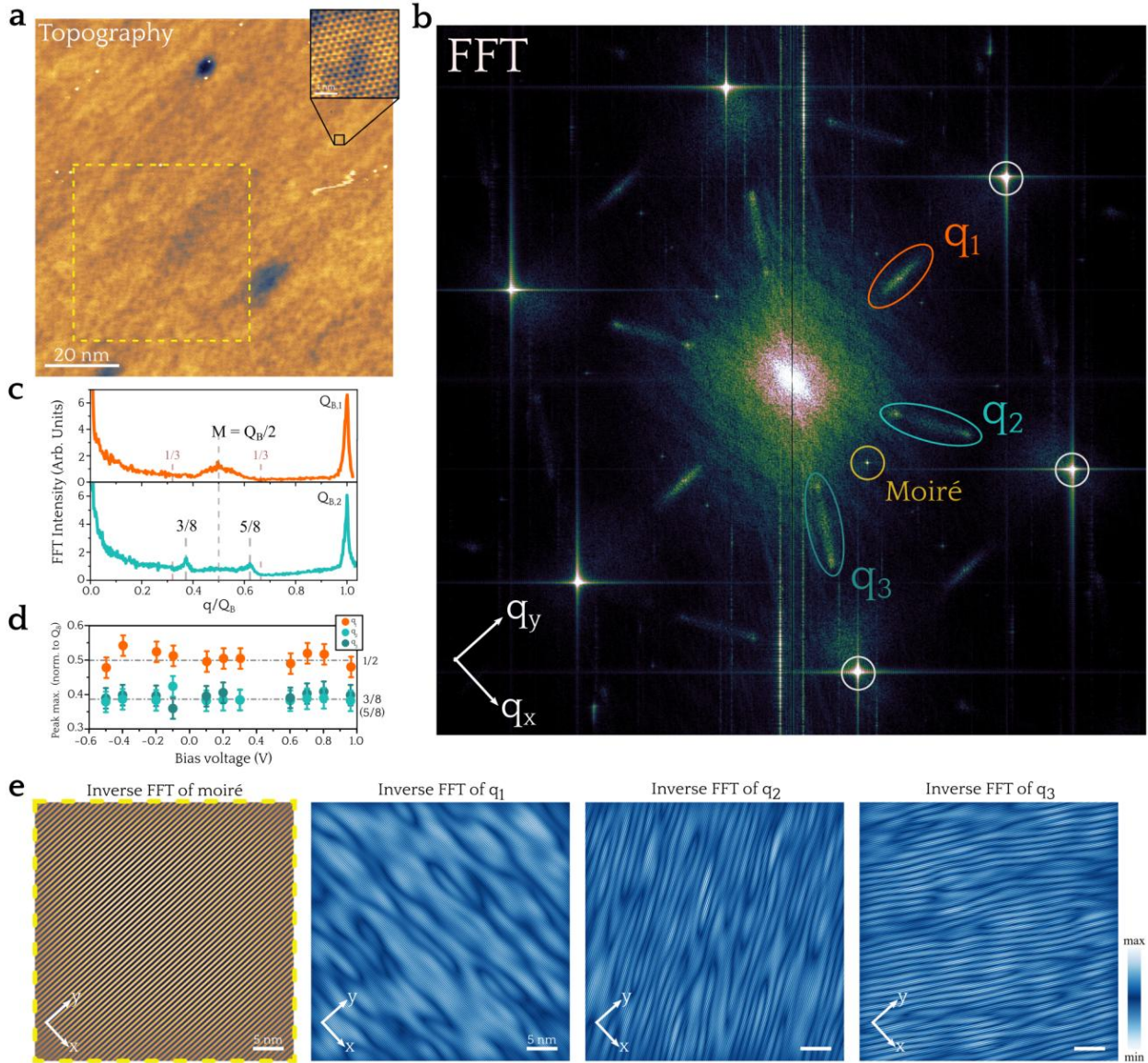

Figure 3. Broken lattice degeneracy in the CDW of $(PbS)_{1.13}TaS_2$. **a**, Large-scale STM image with atomic resolution in 1H-$TaS_2$ ($V_s$ = 0.6 V, $I$ = 500 pA, $T$ = 4.2 K) and **b**, its corresponding Fourier transform (FFT). **c**, Intensity profiles of the FFT along $\Gamma$-$Q_{B,1}$ (orange) and $\Gamma$-$Q_{B,2}$ (cyan). **d**, Energy position of the maxima of the CDW features in the FFT as a function of bias voltage. **e**, Inverse FFT of the moiré peak (left panel) and the $q_i$ elongated features (blue).

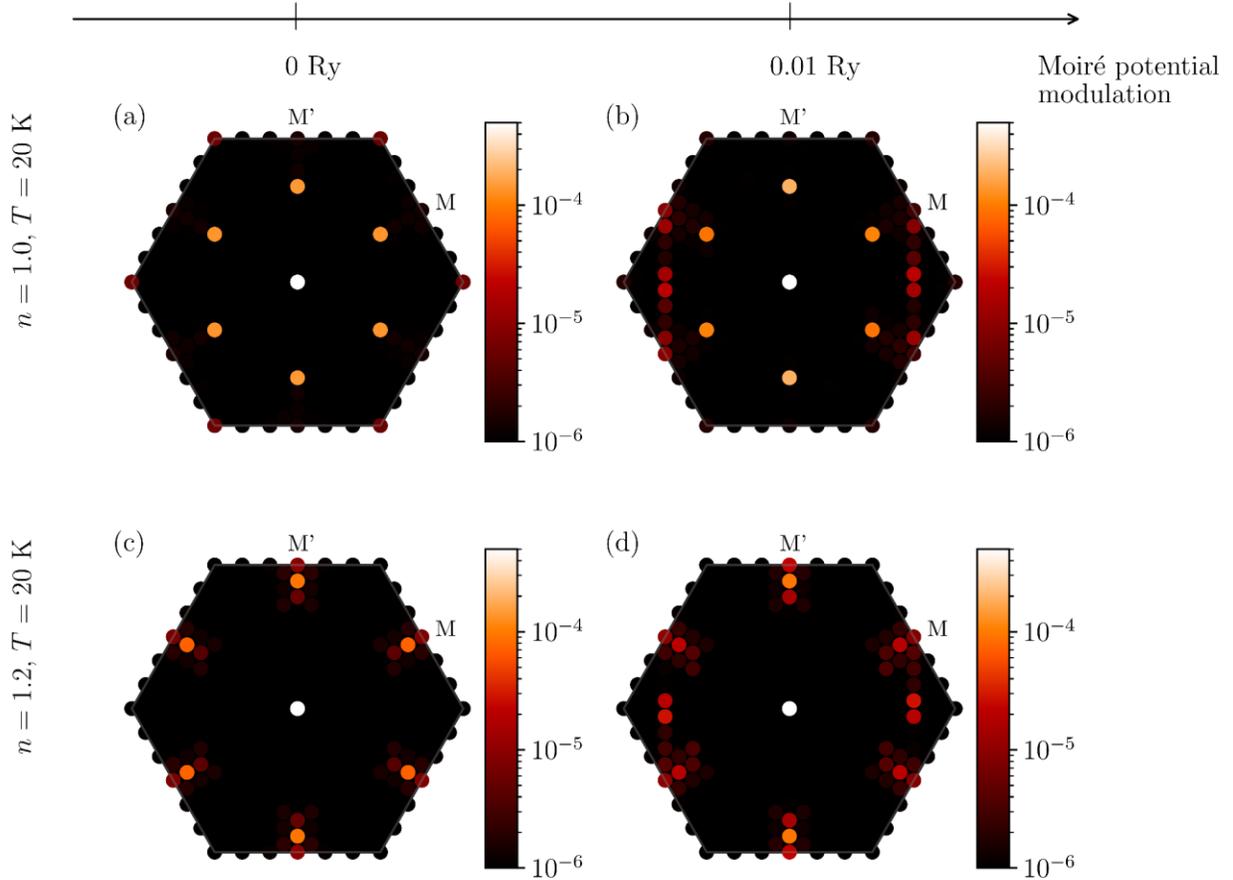

Figure 4. **Influence of charge transfer and moiré potential on the CDW state.** Structure factors from REMD simulations for the replica at $T$ = 20 K for an 18×18 1H-TaS$_2$ supercell. Charge-neutral ($n$ = 1.0; **a,b**) and electron-doped 1H-TaS$_2$ (+0.2e per TaS$_2$ unit cell, $n$ = 1.2; **c,d**) are shown without (**a,c**) and with (**b,d**) a 0.136 meV moiré potential modulation. The undoped supercell **a** verifies the expected 3×3 CDW, while electron doping shifts the system into a (2×2)-like ICCDW configuration in **c**, in line with the short CDW coherence lengths in the experiment. The simulation of interlayer hybridization via a vertically striped electronic potential acting on the downfolded Hamiltonian enforces symmetry breaking as well as a smeared out Moiré contribution around $q_{moiré}$ in the undoped system **b**. Combining the doping-induced shift towards M as well as hybridization effects in **d**, the key experimental CDW signatures are reproduced.

*Acknowledgements*

M.M.U. acknowledges support by the ERC Starting grant LINKSPM (Grant #758558). M.M.U and S.B. acknowledge the Gipuzkoa Quantum 2025 grants (refs: 2025-QUAN-000009-01 and 2025-QUAN-000013-01). M.M.U. and M.N.G. acknowledge support by the grant PID2023-153277NB-I00 funded by MCIN/AEI/10.13039/501100011033. M.N.G is also supported by the Ramon y Cajal Grant RYC2021-031639-I funded by the same institution and EU NextGenerationEU/PRTR. H. G. acknowledges funding from the EU NextGenerationEU/PRTR-C17.I1, as well as by the IKUR Strategy under the collaboration agreement between Ikerbasque Foundation and DIPC on behalf of the Dept. Education of the Basque Government. This work is supported by the Cluster of Excellence "CUI: Advanced Imaging of Matter" of the Deutsche Forschungsgemeinschaft (DFG) - EXC 2056 - project ID 390715994. L.P. and T.W. thank the DFG for funding through the research unit QUAST FOR 5249 (project No. 449872909; project P5). L.P., C.P. and T.W. gratefully acknowledge the computing time made available to them on the high-performance computer "Lise" at the NHR Center NHR@ZIB under the project hhp00063. This center is jointly supported by the Federal Ministry of Education and Research and the state governments participating in the NHR ([www.nhr-verein.de/unsere-partner](www.nhr-verein.de/unsere-partner)). M.R. acknowledges funding by the Deutsche Forschungsgemeinschaft (DFG, German Research Foundation) – Project-ID 555467911 – CRC 1772 / TP A06. S.S. acknowledges enrolment in the doctorate programme Physics of Nanostructures and Advanced Materials from the Advanced Polymers and Materials, Physics, Chemistry and Technology Department of UPV/EHU.

*Author contributions*

M.M.U. conceived the project. T.A. carried out the sample growth and bulk crystal characterization under the supervision of R.P.S.. S.S. and H.G. measured the STM/STS data with the help S.D.. S.S. and P.V.S. measured the STM/nc-AFM data at $T$ = 8 K under the supervision of S.B.. S.S., M.N.G. and M.M.U. analyzed the STM/STS data. L.P. performed the downfolding and MD calculations, which were discussed and interpreted in cooperation with C.P. and T.W.. L.P. and C.P. created figures 4, S10 and S11. M.M.U., T.W., L.P. and C.P. wrote the manuscript. All authors contributed to the scientific discussion and manuscript revisions.

*Competing interests:* The authors declare no competing interests.

Supplementary materials for

# Moiré-induced symmetry breaking of charge order in van der Waals heterostructures

Sandra Sajan, Laura Pätzold, Tarushi Agarwal, Clara Pfister, Haojie Guo, Sisheng Duan, P. V. Sruthibhai, Mariana Rossi, Maria N. Gastiasoro, Sara Barja, Ravi P. Singh, Tim Wehling[*] and Miguel M. Ugeda[*]

[*] *Corresponding authors:* *tim.wehling@uni-hamburg.de* and *mmugeda@dipc.org*

Supplementary note 1. Mesoscopic characterization of bulk $(PbS)_{1.13}TaS_2$ and $(SnS)_{1.15}TaS_2$

The X-ray diffraction (XRD) patterns, recorded at room temperature, of the grown single crystals of $(PbS)_{1.13}TaS_2$ and $(SnS)_{1.15}TaS_2$ are shown in figs. S1a and S1b, respectively. The corresponding optical image of the crystals are presented in the insets.

Bulk superconductivity in $(PbS)_{1.13}TaS_2$ was verified by the appearance of a diamagnetic response in temperature-dependent magnetization measurements carried out under zero-field-cooled warming (ZFCW) and field-cooled cooling (FCC) conditions at $H^\perp$ = 1 mT, displayed in fig. S2a. The onset superconducting transition was measured at $T_C \simeq$ 3.1 K.

Similarly, the bulk superconductivity in $(SnS)_{1.15}TaS_2$ was confirmed by diamagnetic signal in the temperature-dependent magnetization measurement, performed in zero field cooled warming (ZFCW) and field cooled cooling (FCC) mode at $H^\perp$ = 1 mT, displayed in fig. S2b. The onset of the superconducting transition occurs at $T_C \simeq$ 2.9 K. To determine the upper critical field ($H^\perp_{C2}$) value, temperature-dependent resistivity measurements were carried out under various applied magnetic fields oriented along the c-axis (H ∥ c), as shown in fig. S3a. The $H^\perp_{C2}$ value were estimated using the criterion $\rho$ = 0.5·$\rho_n$, where $\rho_n$ indicates the normal state resistivity (at T = 4 K). For H ∥ c orientation, the obtained T dependence of $H^\perp_{C2}$ follows Ginzburg-Landau (GL) model, yielding $H^\perp_{C2}(0) \simeq$ 0.33 T, as shown in fig. S3b. This value is consistent with STM/STS data (see supplementary note 5).

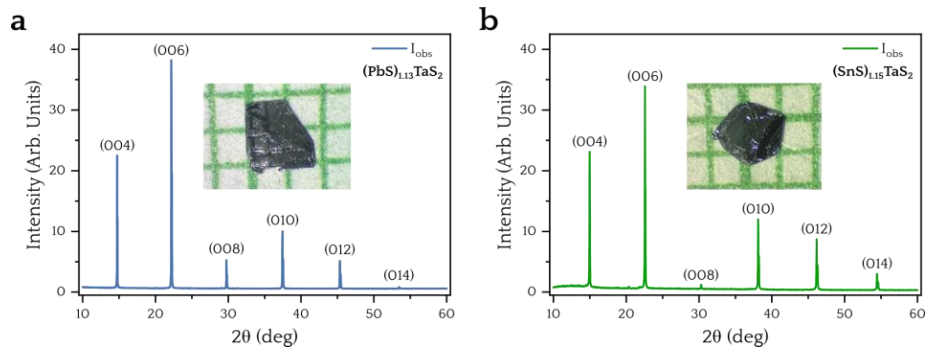

**Figure S1.** Room temperature single-crystal x-ray diffraction patterns of **a**, $(PbS)_{1.13}TaS_2$ and **b**, $(SnS)_{1.15}TaS_2$. The diffraction peaks correspond exclusively to reflections from the basal (00l) planes, confirming the high crystallinity and preferred orientation of the layered crystals along the c-axis. Insets show optical microscopic images of the obtained single crystals.

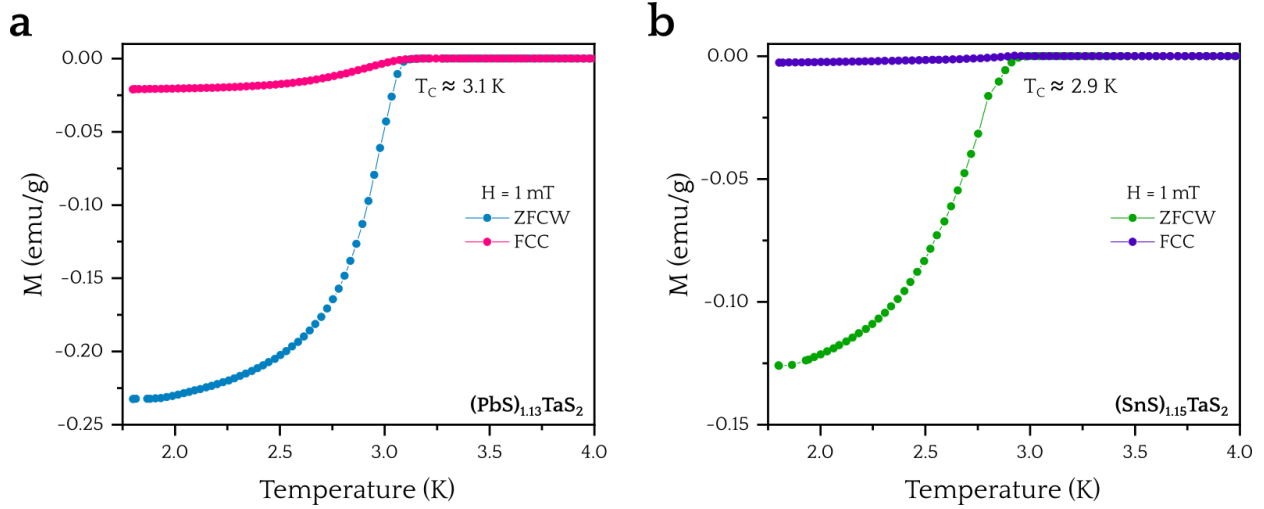

**Figure S2.** Temperature-dependent magnetic susceptibility of (a) $(PbS)_{1.13}TaS_2$ and (b) $(SnS)_{1.15}TaS_2$, showing the onset of diamagnetic shielding associated with the superconducting transition in both systems.

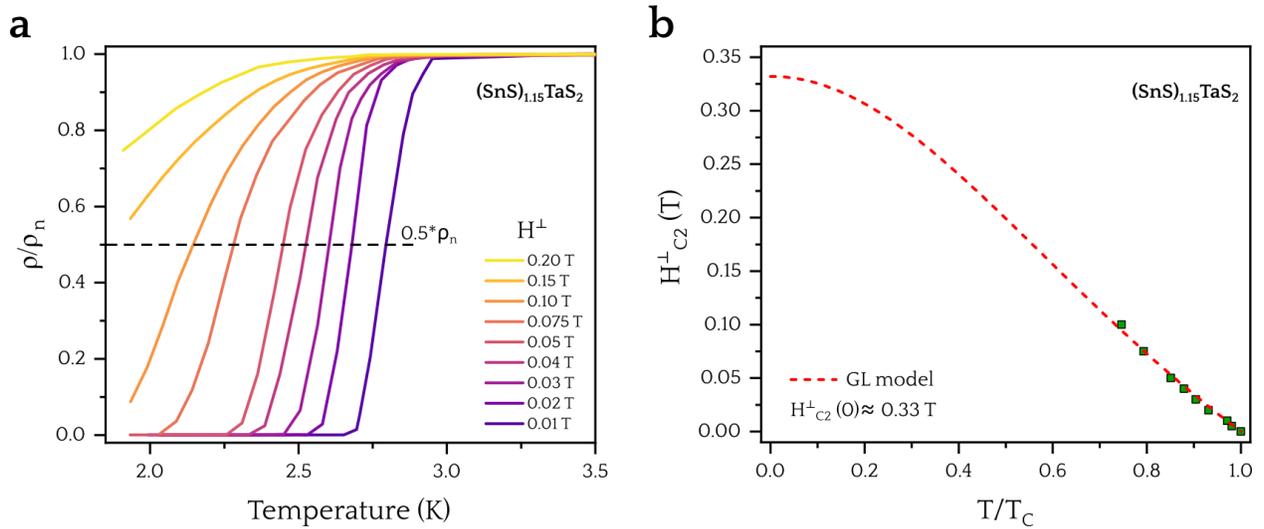

**Figure S3. a**, Temperature-dependent electrical resistivity of $(SnS)_{1.15}TaS_2$ measured under different magnetic fields for $H^{||c}$ orientation ($H^\perp$). **b**, $H^\perp_{C2}$ variation as a function of temperature. The dashed red line indicates the fitting with the GL model.

Supplementary note 2. Atomic scale morphology of the $(SnS)_{1.15}TaS_2$ compound

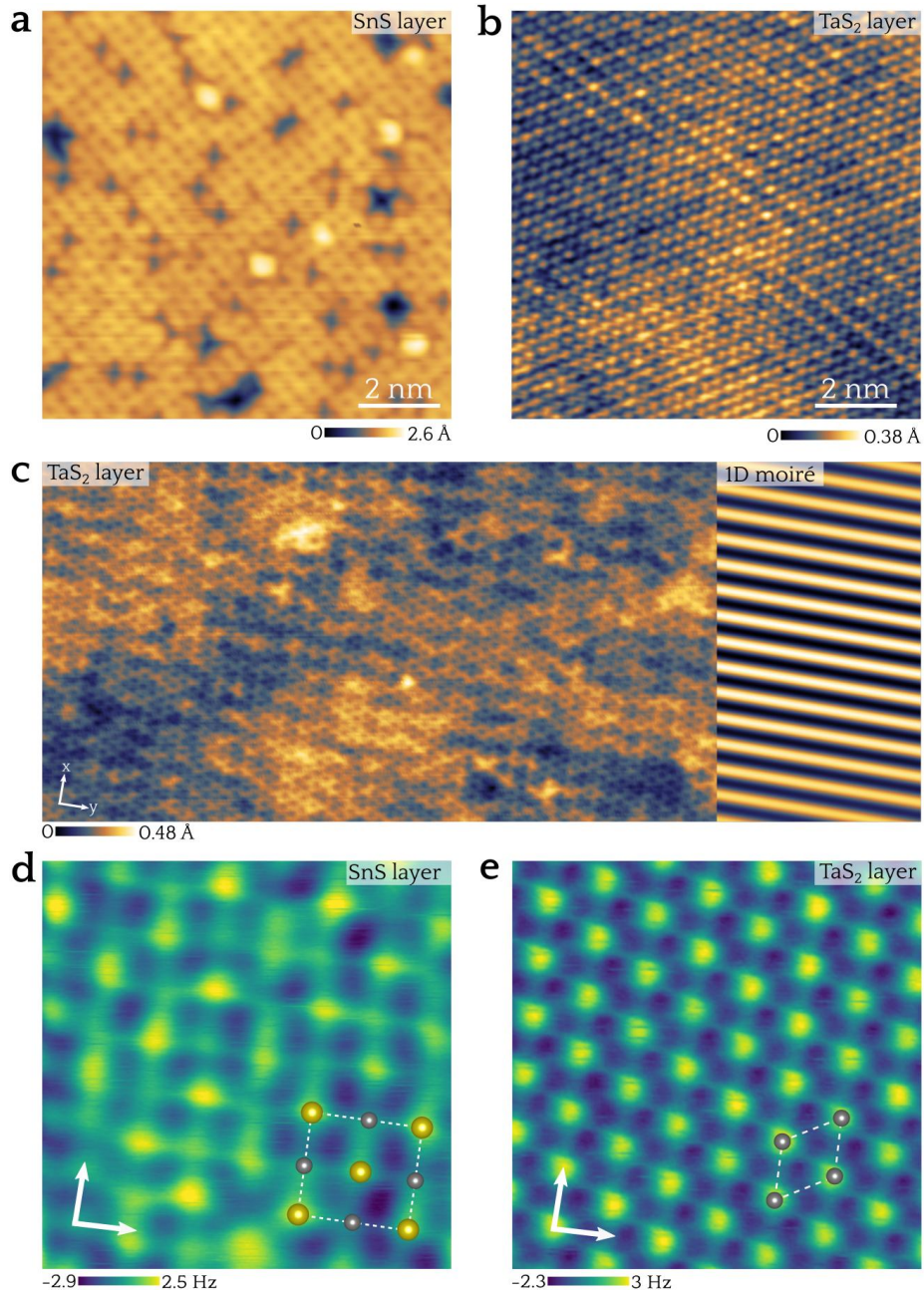

**Figure S4. Atomic scale morphology of $(SnS)_{1.15}TaS_2$.** Atomically resolved STM images of **a**, the SnS layer ($V_s$ = 1.2 V, $I$ = 30 pA, $T$ = 4.2 K) **b**, the 1H-TaS$_2$ layer ($V_s$ = 0.2 V, $I$ = 0.2 nA, $T$ = 3.8 K) and **c**, the 1H-TaS$_2$ layer where the 1D is directly visible in the topography ($V_s$ = -0.2 V, $I$ = 0.15 nA, $T$ = 3.8 K). The right part shows the inverted FFT of the image to guide the visualization of the moiré. **d** and **e**, Constant height nc-AFM images (2.2 nm × 2.2 nm) of the SnS and TaS$_2$ layers, respectively ($T$ = 3.8 K). Yellow (gray) balls correspond to Sn (S) atoms.

Supplementary note 3. Electronic structure of the $(SnS)_{1.15}TaS_2$ compound

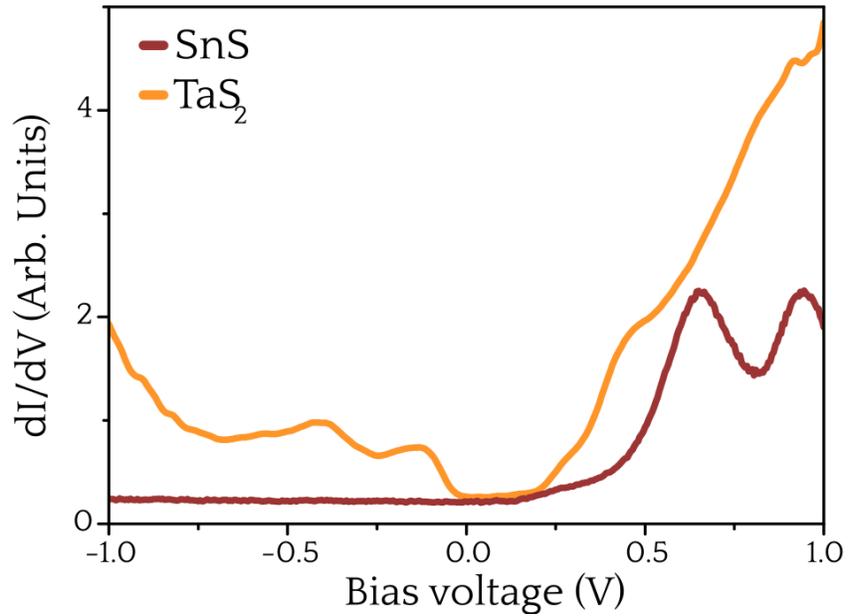

**Figure S5. Large-scale STS on the $(SnS)_{1.15}TaS_2$ compound.** Wide-bias d$I$/d$V$ spectra acquired on TaS$_2$ (orange) and PbS (red) layers ($T$ = 3.8 K).

Supplementary note 4. Superconducting gap fitting of the monochalcogenide layers

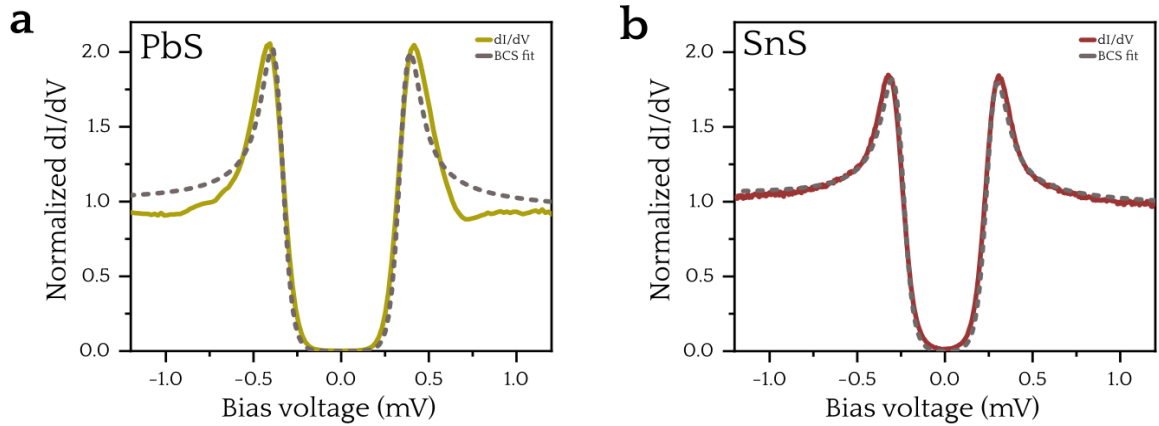

**Figure S6. BCS fit to the superconducting gap of the monochalcogenide layers.** Low-bias d$I$/d$V$ spectra acquired on PbS (a) and SnS (b) measured at $T$ = 0.34 K with the corresponding BCS fits.

Supplementary note 5. Magnetic field dependence of the superconducting gap of $(PbS)_{1.13}TaS_2$

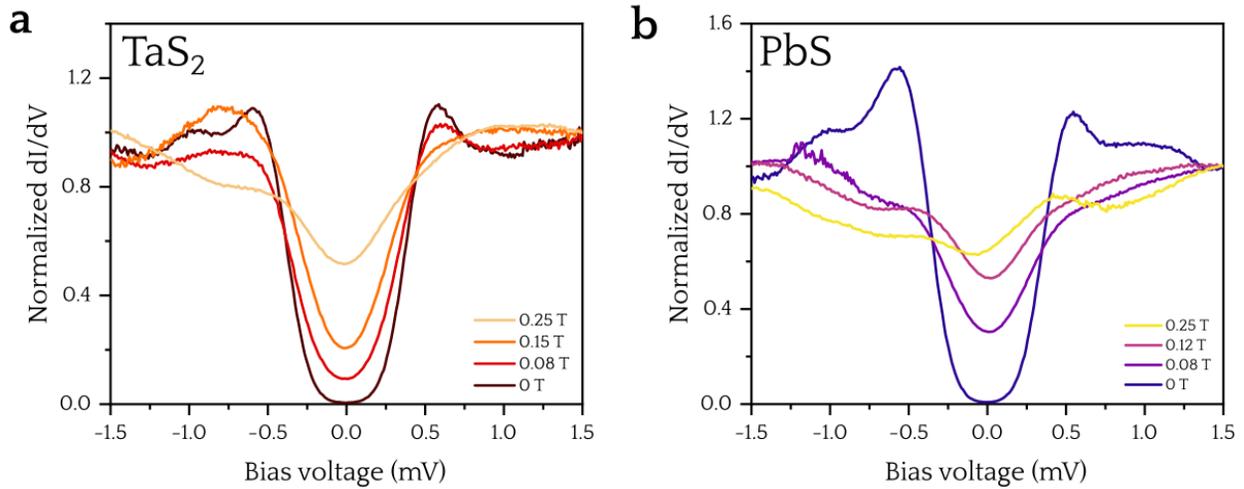

**Figure S7. Magnetic field dependence of the superconducting gap.** Low-bias d$I$/d$V$ spectra sets acquired on TaS$_2$ (**a**) and PbS (**b**) for different out-of-plane magnetic field values ($H^\perp$) measured at $T$ = 0.34 K.

Supplementary note 6. Spatially resolved superconducting gap

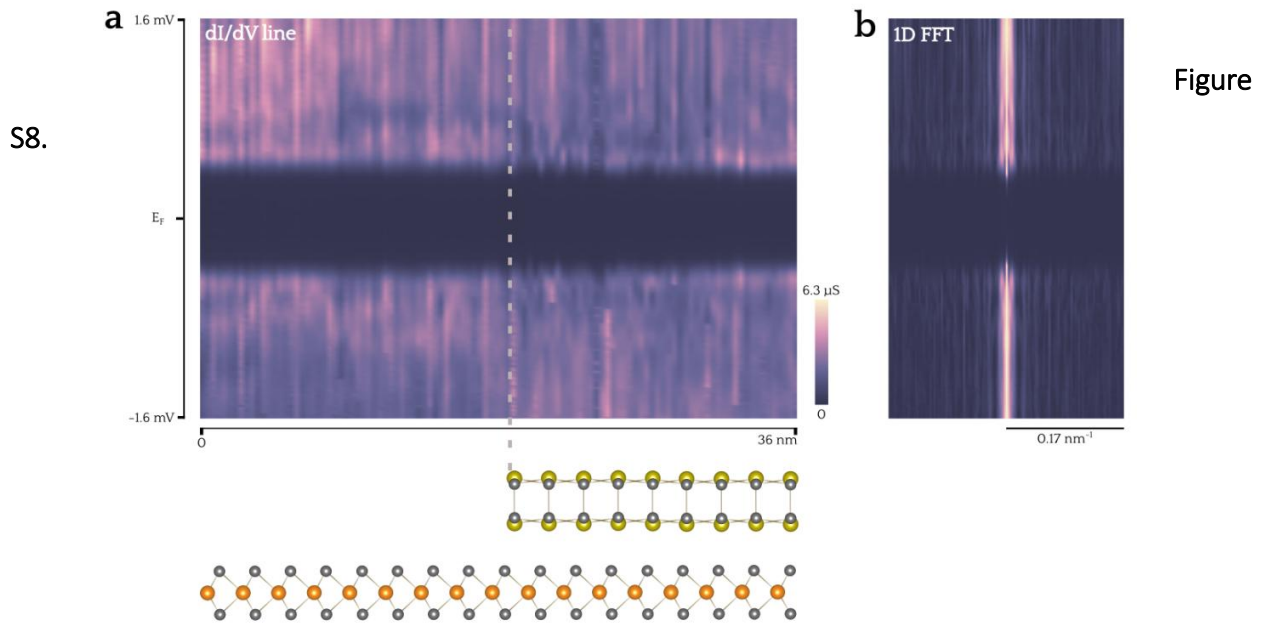

Figure S8. **Spatially resolved low-energy STS. a.** Low-bias d$I$/d$V$ line spectra acquired on adjacent TaS$_2$ (left) and PbS (right) terraces ($T$ = 0.34 K). **b.** Corresponding 1D FFT.

Supplementary note 7. Broken lattice degeneracy in the CDW of (SnS)$_{1.15}$TaS$_2$

Figure S9a shows a large-scale STM image of the TaS$_2$ layer in (SnS)$_{1.15}$TaS$_2$ with atomic resolution, whose FFT is shown in fig.S9b. As in (PbS)$_{1.13}$TaS$_2$, the FFT shows elongated CDW features $q_i$ along the Bragg directions with two distinct internal structures. The moiré peak is also visible at $\mathbf{q}_{moiré} = 0.80$ Å$^{-1}\cdot\mathbf{q}_x$. The $q_1$ feature, orthogonal to $\mathbf{q}_{moiré}$, exhibits intensity widely extended over $\Delta q_{1,FWHM} \simeq 0.42$ Å$^{-1}$ around a maximum at $\frac{1}{2}Q_B$ (2×2 in real space). The $q_2$ and $q_3$ features, rotated by 30° with respect to $\mathbf{q}_{moiré}$, show equal intensities larger than that of $q_1$, which span long arcs of $\Delta q_{2,3,FWHM} \simeq 0.65$ Å$^{-1}$. Unlike $q_2$ and $q_3$ in (PbS)$_{1.13}$TaS$_2$, these features show a roughly constant intensity over their entire extent. As in (PbS)$_{1.13}$TaS$_2$, no intensity has been found in the $q_i$ features at $\frac{1}{3}Q_B$ (3×3 in real space) for TaS$_2$. These observations indicate that the threefold lattice degeneracy of the CDW in the 1H-TaS$_2$ monolayer is also lifted in (SnS)$_{1.15}$TaS$_2$, as it is described for (PbS)$_{1.13}$TaS$_2$ in the main manuscript.

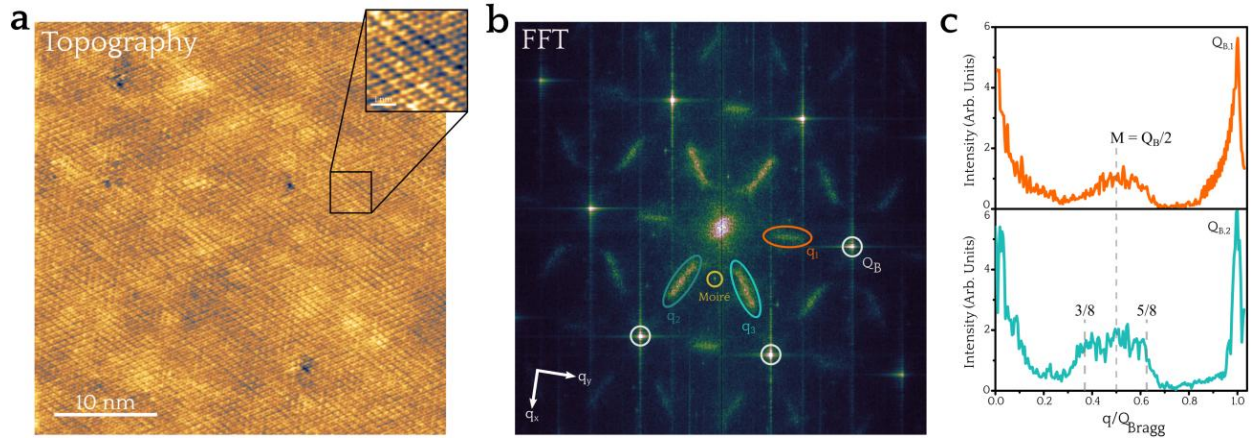

**Figure S9. CDW in the 1H-TaS$_2$ layer of (SnS)$_{1.15}$TaS$_2$.** **a**, Large-scale STM image with atomic resolution in 1H-TaS$_2$ ($V_s$ = 0.08 V, $I$ = 800 pA, $T$ = 3.8 K) and **b**, its corresponding Fourier transform (FFT). **c**, Intensity profiles of the FFT along (0,0) and Q$_{B,1}$ (orange) and Q$_{B,2}$ (cyan).

**Supplementary note 8: Downfolding model of undoped 1H-TaS$_2$**

The results of the downfolding model relaxation for undoped monolayer 1H-TaS$_2$ with a lattice constant of 3.34 Å are shown in fig. S10. The low-energy subspace consisting of three energy bands around the Fermi energy was transformed into a Wannier basis starting from orbitals of Ta $d_{z^2}$, $d_{x^2-y^2}$, and $d_{xy}$ character. The top view of the 18×18 supercell in fig. S10a indicates the relaxation into the native 3×3 CDW reconstruction via arrows, rescaled by a factor of 20 for better visibility. The distortion is limited to the $xy$ direction, with the $z$ direction being undistorted, see the side view in fig. S10a. The reconstruction into the CDW is accompanied by an asymmetric double-well potential in the total energy, as seen in fig. S10b. The global minimum located at a displacement of α = 1 corresponds to the CDW distortion shown, with α = 0 being the structure in the symmetric phase and the displacement being linearly interpolated in-between. By relaxing into the CDW, an energy of about 750 meV is gained in the 18×18 supercell, which equals a normalized binding energy of about 2.3 meV per unit cell.

To visualize the phonon contribution responsible for the CDW instability, we calculate the static structure factor

$$S(\boldsymbol{q}) = \frac{1}{N_{at}^2} \left| \sum_{l=1}^{N_{at}} e^{-i\boldsymbol{q}\cdot\boldsymbol{R}_l} \right|^2$$

see fig. S10 (c). It represents the Fourier transform of the relaxed 3×3 atomic superstructure in fig. S10 (a), while also providing an estimation of the momentum-dependent scattering intensity measured in experiment. The structure factor is peaked in reciprocal space around $\frac{2}{3}\Gamma M$ ($\frac{1}{3}Q_B$), which corresponds to the expected 3×3 CDW reconstruction. The less intense peak at q = K is also characteristically visible for the 3×3 CDW. A closer look at the corresponding DFPT phonon dispersion of the 1H-TaS$_2$ unit cell in fig. S10d also verifies the position of the expected soft phonon mode at $\frac{2}{3}\Gamma M$ ($\frac{1}{3}Q_B$) for the 3×3 superstructure. While no phonon instability can be found at the K point, the corresponding displacements of √3×√3 are commensurate with the 3×3 CDW and couple anharmonically to the unstable modes at $\frac{2}{3}\Gamma M$.

The band structure of the CDW phase (orange) in undoped 1H-TaS$_2$ is shown in fig.S10e, with the weights of the unfolded bands visualized via logarithmic color bar and markersize. The bands and the Fermi level of the symmetric phase are shown in blue. The band structure of the CDW phase is unfolded from the supercell to the unit cell, and the Fermi level (indicated via dashed lines) stays nearly constant between the symmetric phase and the CDW. Additionally, new band contributions located at $\frac{2}{3}\Gamma M$ are found in the band structure of the CDW phase, in line with the wavevector of the soft phonon mode found in the phonon dispersion in fig.S10d.

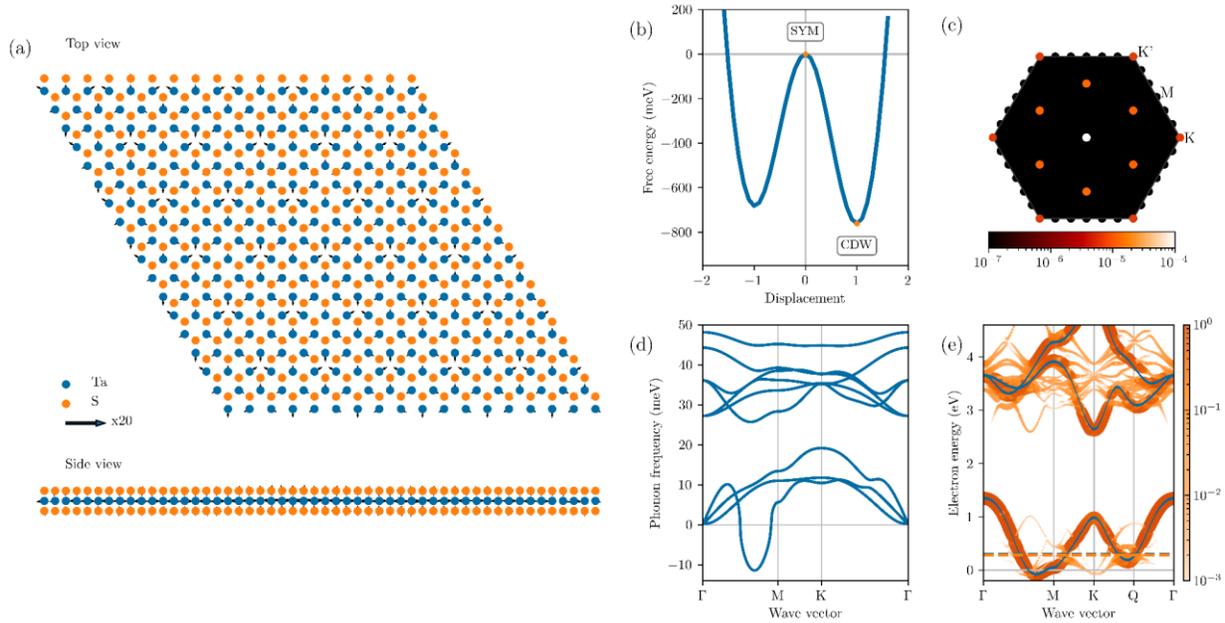

**Figure S10. Model relaxation of undoped 1H-TaS$_2$.** 3×3 CDW in undoped monolayer 1H-TaS$_2$. **a**, Top and side view of the relaxed crystal structure from model calculations performed on an 18×18 supercell with a lattice constant of 3.34 Å. The relaxation into the 3×3 charge density wave configuration is indicated via arrows that are rescaled by a factor of 20. **b**, Total free energy as function of the displacement α for the 3×3 CDW shown in (a). The relaxation started from the symmetric phase (SYM, α = 0), revealing a minimum at α = 1 corresponding to the CDW distortion. The resulting total energy gain of about 750 meV represents the formation energy of the CDW. **c**, Structure factor $S(\mathbf{q})$ of the relaxed structure in (a), with the intensity peaks at $\frac{2}{3}\Gamma M$ of the phononic Brillouin zone indicating the 3×3 CDW. A less intense contribution at the K and K' points, also characteristic for the 3×3 CDW, can be seen. **d**, Phonon dispersion of 1H-TaS$_2$ obtained via DFPT. The soft mode at $\frac{2}{3}\Gamma M$ is indicative of the 3×3 CDW. **e**, Reconstructed band structure of the low-energy subspace as obtained by DFT (blue) and via model relaxation into the CDW structure

(orange). The logarithmic color bar (right) as well as the size of the markers are used to visualize the weights of the unfolded bands. The relaxed band structure of the CDW has been unfolded from the supercell to the unit cell, revealing several replica bands, associated with the symmetry breaking of the CDW phase. The nearly identical Fermi energies of both band structures are indicated via dashed lines in blue and orange, respectively.

**Supplementary note 9: Influence of electron doping on CDW symmetry in 1H-TaS$_2$**

Analyzing the influence of electron doping on the 1H-TaS$_2$ monolayer in our downfolding model reveals a variety of possible CDW symmetries that our model relaxes into for different doping levels of the unit cell. The most commonly appearing CDW configurations were the commensurate 3×3 CDW for the undoped system and the commensurate 2×2 CDW for higher electron doping, with an incommensurate CDW (wavevector of $\frac{8}{9}$ΓM, ~2.25×2.25) in direct energetic competition.

Via the difference in corresponding binding energies between the energetically most favorable and the second most favorable (IC)CDW, we visualize the occurrence of different configurations as a function of the doping and the electronic smearing in a CDW phase diagram in fig. S11. For each doping, an ab initio based model was set up, with the binding energies of each (IC)CDW being calculated via enforcing the input structure for model calculations at values of different doping and smearing (see grey circles). The rest of the energy landscape is interpolated between these values. For each (IC)CDW symmetry, the regions where it is the energetic ground state are indicated via the shading of the colorbars, with different (IC)CDW symmetries being energetically favorable for different doping levels. Note that the colorbars on the right are scaled for best visibility and do not share a unified maximum. The undoped system favors the 3×3 symmetry (pink), with higher electron doping it undergoes a shift towards the 2×2 CDW as the energetic ground state (orange). For even higher electron doping, the 2×2 CDW competes with the 2×2-like ICCDW (green), which becomes slightly favorable. Between these regions, white phase boundaries are visible where both (or multiple) CDW configurations show identical binding energies. For higher electronic smearing (above ≈ 60 meV) as well as for higher doping ($n > 1.25$), no CDW configuration is found to be stable. We therefore located a region in the phase diagram where dominant 2×2 contributions are expected for an electron doping of approximately $n$ = 1.1 to $n$ = 1.2, therefore doping shifts the dominant structure factor contribution towards the M point.

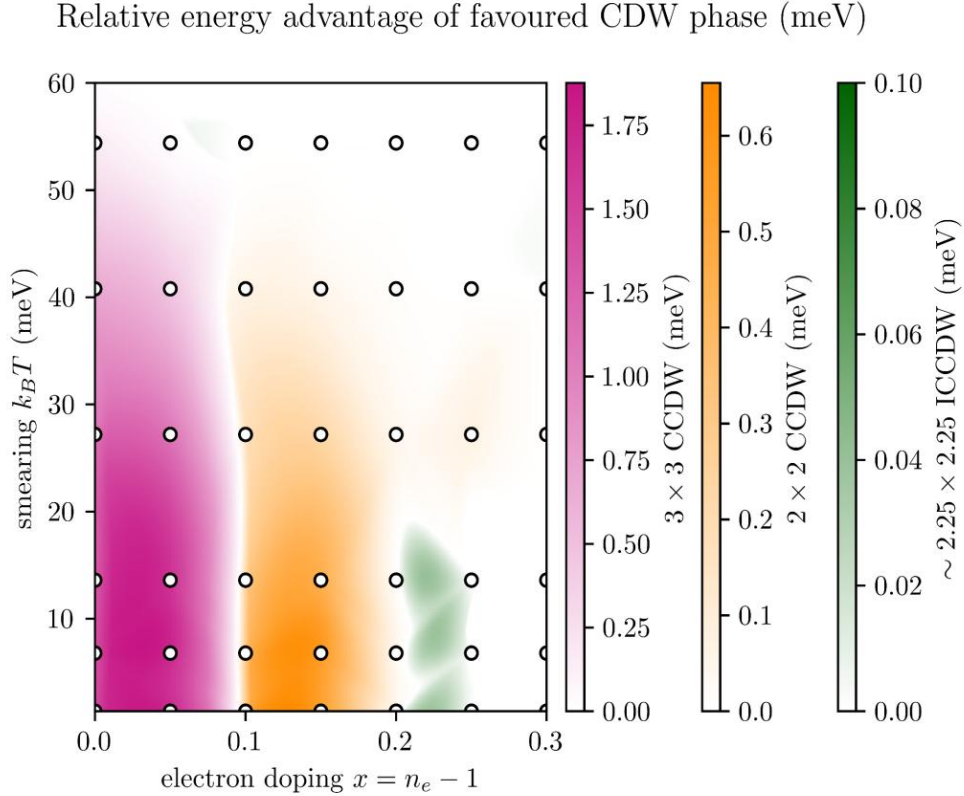

**Figure S11. CDW phase diagram of electron-doped 1H-TaS₂.** Phase diagram of monolayer 1H-TaS$_2$ as a function of the electronic smearing $k_BT$ and the electron doping $x$. The binding energies of each (IC)CDW were calculated via enforcing the input structure for model calculations at different doping and smearing levels (white circles), with the rest of the energy landscape being interpolated between. The differently colored regions show the expected (IC)CDW ground state by indicating the energy difference between the most favorable and the second most favorable configuration. The 3×3 CDW shown in pink is dominant for the undoped system, while the 2×2 CDW (orange) equals the ground state for an electron doping of $n$ = 1.1-1.2. For higher doping, an ICCDW with a wavevector of $\frac{8}{9}\Gamma M$ (green) emerges. Note that the colorbars on the right are scaled for best visibility. Between these regions, white phase boundaries are visible where both (or multiple) CDW configurations show identical binding energies. For electronic smearing above 60 meV as well as for doping above $n$ = 1.25, no CDW configuration is found to be stable.